\documentclass[showpacs,preprintnumbers,amsmath,amssymb,prl]{revtex4}

\usepackage{graphicx}
\usepackage{dcolumn}
\usepackage{bm}

\begin{document}

\title{Comment on "Theory of helimagnons in itinerant quantum systems" by D.Belitz, T.K.Kirpatrick and A.Rosch and "Cubic magnets with Dzyaloshinskii-Moriya interaction at low temperatures" by S.V.Maleyev}

\author{S. V. Maleyev}
\affiliation{Petersburg Nuclear Physics Institute, Gatchina, St.\ Petersburg 188300, Russia}
\email{maleyev@sm8283.spb.edu}




\maketitle

 In  recent papers \cite{B} and \cite{M}  low-energy spin-wave excitations in non-centrosymmetric cubic helimagnets  were   studied theoretically.  In \cite{B} was claimed that these spin-waves are gapless Goldstone excitations due to translation invariance of the spin structure along the helix axis $\bf k$ which can be broken  by spin-lattice interaction only. Corresponding spin-wave gap has to be negligibly small.  However in \cite{M} the spin-wave gap $\Delta$ was calculated in the Hartree-Fock approximation (HFA) as a result of interaction of the long-wave and short-wave excitations which momenta are of order of the inverse  lattice constant $1/a$. So in continuum approximation used in \cite{B}  one has to obtain ultra violet divergence. Hence the lattice structure can not be ignored. As a result  explicit form of the Dzyaloshinskii interaction (DI) plays crucial role.

  In \cite{B} this interaction has the form proposed in \cite{D} 
\begin{equation}
H_{D,I}=D\int d^3r\mathbf{S(r)}\cdot[\nabla\times \mathbf{S(r)}],
\end{equation}
where  $\mathbf{S(r)}$ is a spin density. In reality the DI acts between different spins. So it feels the lattice structure. Corresponding expression used in \cite{M} is given by 
\begin{equation}
	H_{D,II}=(1/2)\sum
	D_{\mathbf{R,R'}}(\nabla_{\mathbf R}-\nabla_{\mathbf R'})\mathbf{[S_{\bf} \times S_{\mathbf R'}]},
\end{equation}
where $\mathbf {S_{R}}$ is a spin at the lattice point $\mathbf R$ and $D_{\mathbf{R,R'}}=D_{\mathbf{R',R}}$ is strength of the DI. In this expression the Dzyaloshinskii vector $\vec d_{\mathbf{R,R'}}=D_{\mathbf{R,R'}}(\nabla_{\mathbf{R}}-\nabla_{\mathbf R'})$ has correct symmetry $\vec d_{\bf{R,R'}}=-\vec d_{\bf{R',R}}$ and $\vec d_{\bf{R=R'}}=0$ due to the vectors $\nabla$. So we do a  natural  assumption  that $ D_{\bf{R=R'}}\neq 0$.  

This minor difference in the form of the DI does not affect the helical spin structure and spin-wave spectrum in the bilinear approximation. Corresponding results obtained in both papers are practically the same. Nevertheless if one considers spins as quantum operators the gap appears in the case of interaction (2) only. The simplest case of zero magnetic field and  planar helix is discussed below while the general case  will be considered elsewhere \cite{MG}.   For the planar helix the lattice spin has the form $\bf {S_R}=S^\zeta_{\bf R} \hat \zeta_{\bf R} +S^\eta_{\bf R} \hat \eta_{\bf R} +S^\xi_{\bf R} \hat \xi_{\bf R}$ where $\hat\zeta_{\mathbf R}=2Re\{\mathbf A e^{i\mathbf{k\cdot R}}\} \quad;
\hat\eta_{\mathbf R} =-2Im\{\mathbf A e^{i\mathbf{k\cdot R}}\};\quad
\hat \xi_{\bf R} =\hat c$,  $\mathbf A=(\hat a-i\hat b)/2$,  and the unit vectors $\hat a$, $\hat b$ and  $\hat c$ form the right-handed frame.  The spin operators are given by the well known expressions: $S^\zeta_{\bf R} =S-(a^+a)_{\mathbf R}$, $S^\eta_{\mathbf R} =-i(S/2)^{1/2}[a_{\mathbf R} -a^+_{\mathbf R} -(a^+a^2)_{\mathbf R} /(2S)]$ and $S^\xi_{\mathbf R} =(S/2)^{1/2}[a_{\mathbf R} +a^+_{\mathbf R} -(a^+a^2)_{\mathbf R} /(2S)]$.
 
  In \cite{M} the Hamiltonian is a bilinear form of the spin operators belonging to different lattice points and the the fourth-order spin-wave interaction may be written as  a normal product of operators $a^+$ and $a$ [see Eq.(D3) in \cite{M}]. As a result in the Hartree-Fock approximation the exchange part of interaction leads to the gap [Eqs. (D7) and (D8)]. It is a consequence of anomalous average $f_{\bf q}=<a^+_{\bf q}a^+_{-\bf q}>$  which appears  in the bilinear approximation due to  total spin non-conservation summoned by the DI.
  
   In the traditional form of the DI given by Eq.(1) its fourth order part is given by
\begin{equation}
	V^{(4)}_{D,I}=-D(\mathbf k\cdot \hat c)\sum \{(a^+a)(a^+a)+[(a-a^+)a^+a^2+a^+a^2(a-a^+)]/4\}
\end{equation}
where all operators belong to the same lattice point $\mathbf R$. This expression can be divided on two parts: The first one has form of a normal product  and may be neglected  as in \cite{M}. The second has  a bilinear form $-D(\mathbf k\cdot \hat c)\sum(a^+a-a^2/2)$. This non-Hermitian term  treated as in \cite{M} (see also \cite{P})  cancels the gap produced by the above mentioned exchange contribution. Hence we have not a gap in the case of the DI in the conventional form (1), but it appears  if one takes into account that the DI acts between different spins. The gap appears also if one assumes
commutativity of spin components belonging to two spins  in Eq.(1).

In the helimagnets the three spin-wave interaction appears. In the second order perturbation theory it can give negative correction to the square of the gap which has the same $1/S$ order as the Hartree-Fock part \cite{P},\cite{C}. However in considered case at low momenta $\mathbf q$ this interaction is proportional to $\mathbf{(q\cdot k)}$  [see Eq.(11) and discussion below Eq.(19) in \cite{M} ] with small coefficient. Moreover explicit calculations show that this correction does not contribute to the gap.   

 I wish to note also that the $1/S$ expansion is a standard method in the spin-wave theory. In this respect I can mention well known Oguchi corrections to the spin-wave stiffness in ferro and antiferromagnets and the gap calculated in frustrated antiferromagnets \cite{P}.
 
 In conclusion I have to point out that the gap was considered in  \cite{M} for explanation of the helical system behavior in magnetic field perpendicular to the helix wave vector $\mathbf k$ experimentally investigated in \cite{L},\cite{G}.

   This work is supported in part by the RFBR (projects No 05-02-19889, 06-02-16702 and 07-02-01318) and the Russian State Programs "Quantum Macrophysics" and "Strongly correlated electrons in Semiconductors, Metals, Superconductors and magnetic Materials" and Russian State Program "Neutron research of solids".

\end{document}